\documentclass[sigconf,nonacm]{acmart}

\AtBeginDocument{%
  }

\setcopyright{acmlicensed}
\copyrightyear{2026}
\acmYear{2026}
\acmDOI{XXXXXXX.XXXXXXX}
\acmConference[ICAIF '26]{7th ACM International Conference on AI in Finance}{Nov 14--17,
  2026}{Milan, Italy}
\acmISBN{978-1-4503-XXXX-X/2018/06}

\usepackage{graphicx}
\usepackage{subcaption}
\usepackage{arydshln}
\usepackage{multirow}
\usepackage{amsmath}
\usepackage{booktabs}

\begin{document}

\title{No Data Is Not No Risk: Visibility Aware Graph-Based Inference of Business Conduct Risk}

\author{Tsuyoshi Iwata}
\email{tsuyoshi.iwata@reprisk.com}
\orcid{0009-0000-7579-3920}
\affiliation{%
  \institution{RepRisk AG}
  \city{Zurich}
  \country{Switzerland}}

\author{Johannes Laurmaa}
\email{johannes.laurmaa@reprisk.com}
\orcid{0009-0001-1790-4036}
\affiliation{%
  \institution{RepRisk AG}
  \city{Zurich}
  \country{Switzerland}}

\author{Ryohei Hisano}
\email{hisanor@g.ecc.u-tokyo.ac.jp}
\orcid{0000-0002-4452-0302}
\affiliation{%
 \institution{The University of Tokyo}
 \city{Tokyo}
 \country{Japan}}

\affiliation{%
  \institution{The Canon Institute for Global Studies}
 \city{Tokyo}
 \country{Japan}}


\begin{abstract}
The monitoring of business conduct risk is hindered by sparse, uneven, and visibility-biased data. Prior studies show that business conduct risk information and media coverage propagate through supply chain, peer, and corporate structure networks, yet incident records remain incomplete for many firms. As a result, the absence of reported events could reflect limited coverage rather than the absence of underlying business conduct risk. This paper examines whether inter-firm relationships can improve the prediction of future recorded conduct related incidents, particularly among firms with limited prior visibility. We formulate the task as Positive--Unlabeled node classification on a corporate ownership graph, where firms with recorded incidents are treated as labeled positives and firms without recorded incidents remain unlabeled. We then propose a visibility- and relation-aware GCNII framework that combines relation specific message passing with non-negative Positive--Unlabeled learning to account for positive contamination in the unlabeled set. In a forward-looking evaluation, the proposed approach achieved the strongest observed ranking performance relative to non-graph- and simple graph-based benchmarks. The results further show that graph-based inference retains its predictive value among firms without prior recorded incidents. These findings demonstrate the value of inter-firm relational structure as a complementary source of information for extending risk prioritization beyond directly observed incident histories.
\end{abstract}



\maketitle

\section{Introduction}

The assessment of business conduct risk relies heavily on externally observed incidents, yet such records are sparse and unevenly distributed across firms. Initial data availability can itself influence subsequent monitoring: firms that receive an environmental, social, and governance (ESG) rating attract greater analyst coverage, creating a feedback mechanism in which firms already receiving attention face further scrutiny \citep{wei_esg_2026}. Media coverage is likewise concentrated among larger, more profitable, and more visible firms \citep{he_media_2024}, and the visibility of negative conduct events varies substantially across industries and issue types \citep{derrien_esg_2025}. Evidence from private equity portfolios further shows that many active firms have no recorded negative incidents \citep{abraham_esg_2024}. Consequently, the absence of an observed incident cannot be straightforwardly interpreted as evidence of low underlying risk.

This visibility problem is critical because externally reported conduct information influences both monitoring and economic outcomes. Reported incidents attract analyst attention \citep{he_does_2024}, while ESG-related information concerning suppliers and other connected firms can affect the focal firm \citep{wang_spillover_2025}. More broadly, media scrutiny and reputational pressure influence investor behavior, managerial responses, and subsequent corporate conduct \citep{li_framing_2024,gong_consequences_2026,pak_managers_2026,zhuang_how_2025}. Incident records are therefore informative, but they are also generated through a selective observation process.

Inter-firm relational structure may provide information that is unavailable from firm-level incident histories alone. Prior research has documented the associations and spillovers that exist across supply chain, peer, and corporate structure relationships \citep{ji_data_2025,huang_spillover_2026,wang_spillovers_2026,saeed_parent_2024}. Graph-based models have also been used to represent financial dependencies and perform firm-level classification tasks \citep{tan_direction-sensitive_2026,wu_industry_2023,hisano_prediction_2020}. These findings motivate the use of inter-firm relations as an additional source of information for identifying firms whose observed incident histories may be incomplete.

Hence, we formulate business conduct risk prediction as a Positive--Unlabeled (PU) node classification problem on a directed, multi-relational corporate graph. Firms with incidents recorded during the reference period are treated as labeled positives, whereas firms without recorded incidents are not assumed to be negative but instead remain unlabeled. We then developed HeteroGCNII, a relation-aware extension of GCNII that models relation types and directions through separate message passing channels and is trained using a non-negative PU objective. The model was evaluated prospectively: information available through 2024 was used to rank firms according to whether they recorded at least one conduct related incident during 2025.

The empirical analysis in this study addresses five questions:

\noindent\textbf{RQ1}: Does historical conduct risk coverage vary systematically across firms?

\noindent\textbf{RQ2}: Do graph-based models outperform non-graph baselines in predicting future recorded incidents?

\noindent\textbf{RQ3}: In which observability related segments are the gains from graph-based inference largest?

\noindent\textbf{RQ4}: Which inter-firm relation types contribute most to graph-based prediction?

\noindent\textbf{RQ5}: Does graph-based inference retain predictive value among firms with no previously recorded incidents?

The results show that the historical incident coverage varied significantly across firm size, country, and sector. Graph-based models outperformed random, feature only, and simple graph-based baselines, and HeteroGCNII, when trained using the non-negative PU objective, achieved the strongest overall ranking performance. The predictive lift is generally larger in segments with weaker historical incident coverage, and relation ablation analysis identified current subsidiary and operating unit relationships as the most informative relation channel. The model also produced meaningful forward-looking rankings among firms without prior recorded incidents, indicating that inter-firm relational structure adds predictive value beyond firm-level attributes and historical visibility alone.

This paper makes three main contributions. First, it introduces a graph-based PU formulation for business conduct risk prediction under selective incident observability. Second, it proposes HeteroGCNII, a relation-aware GCNII framework for directed, multi-relational corporate graphs, and evaluates the effect of PU risk correction relative to conventional cross-entropy training. Third, it reports a forward-looking and observability stratified evaluation that examined not only aggregate predictive performance, but also where relational information is most useful and which inter-firm relation types carry the strongest predictive signal.

\section{Background and Data}

\subsection{Observability and PU Setting}

\begin{figure*}[t]
\centering
\begin{subfigure}{0.48\textwidth}
\centering
\includegraphics[width=\linewidth]{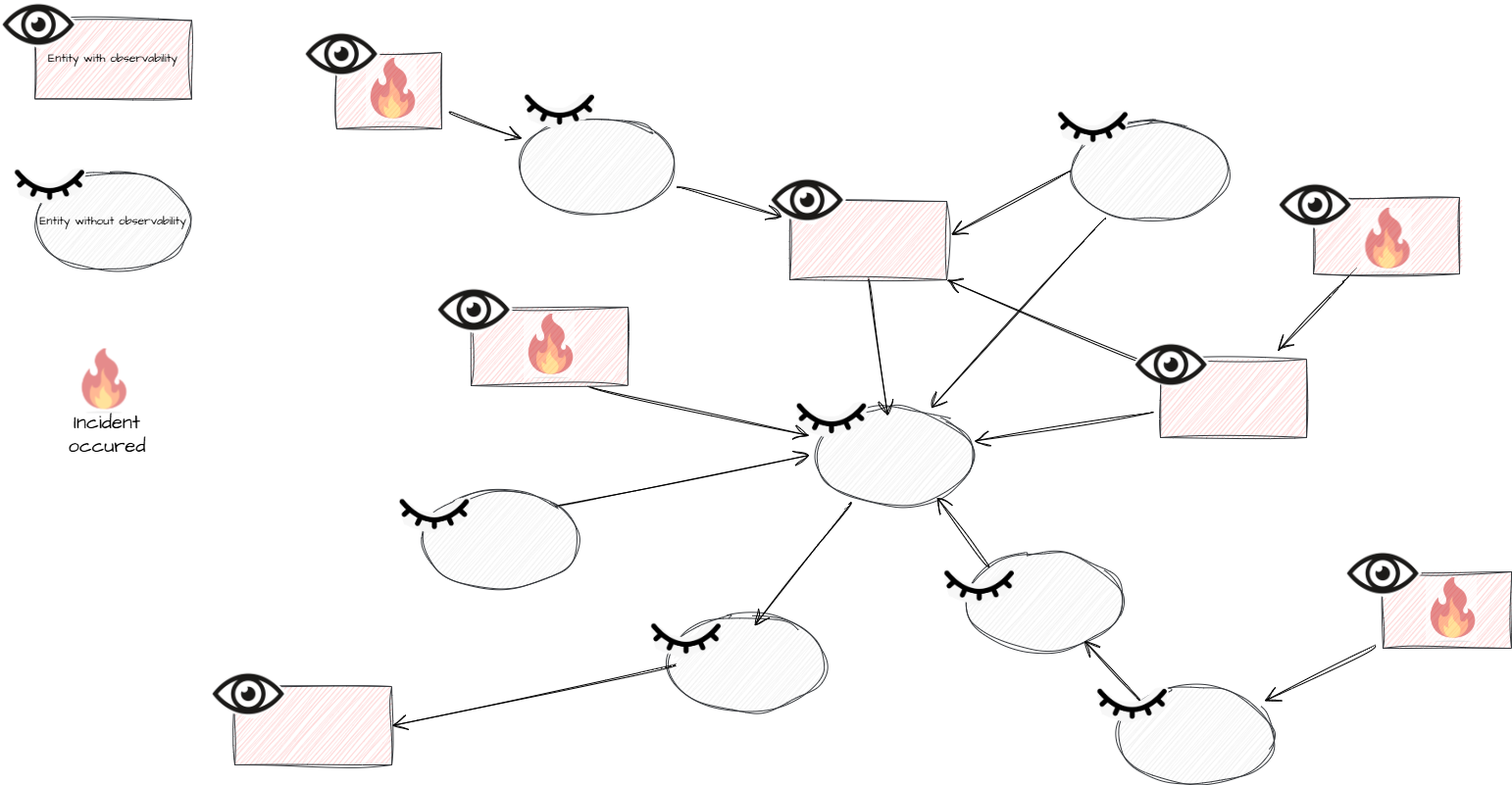}
\caption{Entities with and without recorded incident histories in the inter-firm graph.}
\label{fig:kg}
\end{subfigure}\hfill
\begin{subfigure}{0.48\textwidth}
\centering
\includegraphics[width=\linewidth]{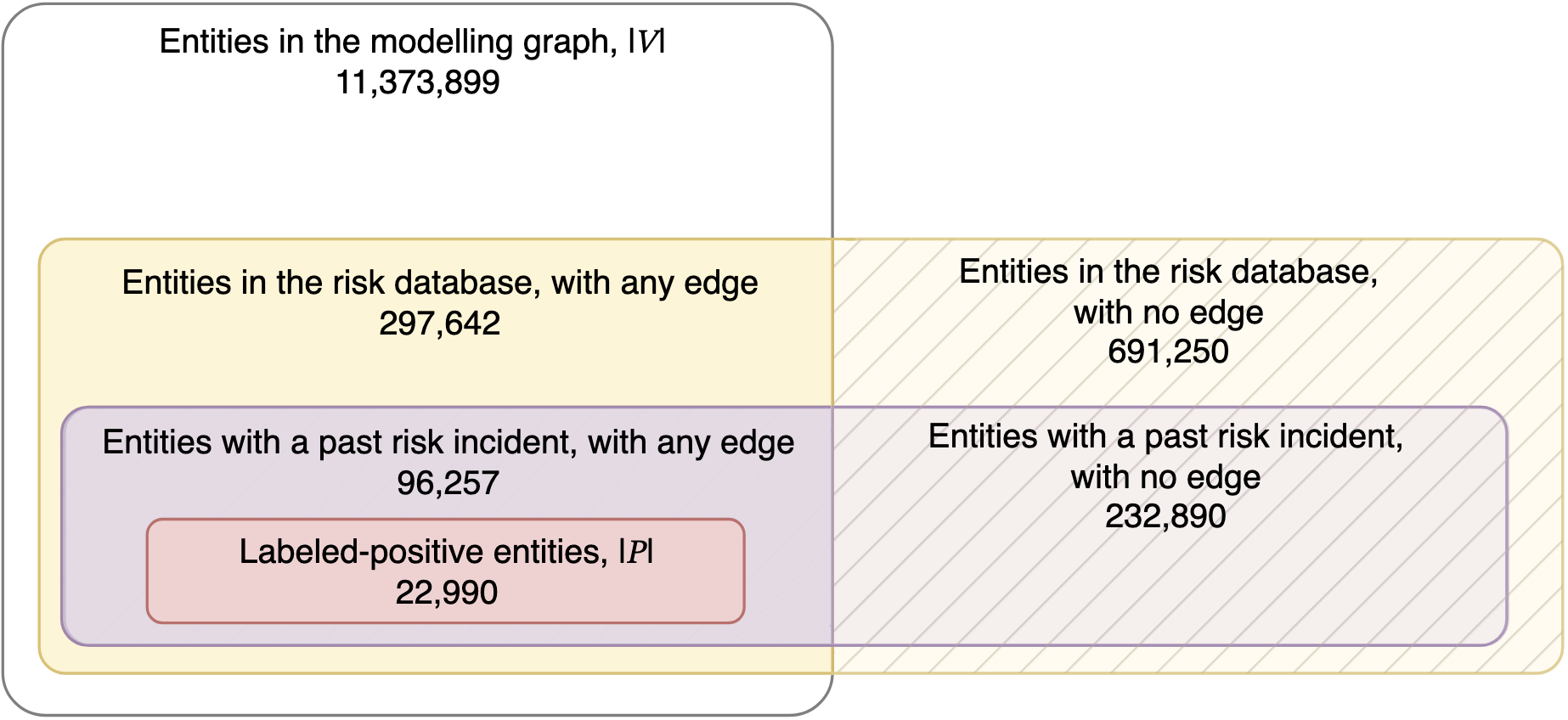}
\caption{Overlap between incident and relational-data coverage.}
\label{fig:coverage}
\end{subfigure}
\caption{``No risk versus no data'' problem. Missing incident records may indicate either low underlying risk or limited observability, while inter-firm relations provide an additional source of information.}
\label{fig:challenge}
\Description{The left panel shows observed and unobserved companies connected through inter-firm relations. The right panel shows the overlap between entities with recorded incidents and entities covered by relational data.}
\end{figure*}

In business conduct risk data, the absence of a recorded incident does not necessarily indicate low underlying risk. Incidents may remain unobserved because of limited media, language,  geographic coverage, or the lower scrutiny applied to small and private firms. Figure~\ref{fig:challenge} illustrates this problem: Figure~\ref{fig:kg} shows entities with and without recorded incident histories in the inter-firm graph, whereas Figure~\ref{fig:coverage} shows the overlap between incident and relational-data coverage.

Only about 33\% of the entities in the proprietary business conduct risk database used in this study have had a recorded business conduct risk signal since 2007, leaving approximately 67\% without an observed incident history. Coverage also varies systematically across company size, country, and sector (Table~\ref{tab:rq3-subgroup}), indicating that missing incident records are not randomly distributed. These patterns motivate the observability analyses for RQ1 and RQ3.


We therefore treat entities with at least one recorded conduct-related incident during the reference period as labeled positives and all remaining entities as unlabeled rather than confirmed negatives. The unlabeled set may contain both low-risk firms and firms whose incidents have not been observed, yielding the PU node-classification setting formalized in Section~\ref{sec:problem-formulation}.

\subsection{Inter-Firm Network}

We represent the corporate universe as a directed, multi-relational graph $G=(V,\{E_r\}_{r\in\mathcal{R}},X)$, where each node $v\in V$ is a legal entity, $E_r$ contains the directed edges associated with relation type $r$, and $X$ contains the node attributes. The entities were drawn from a proprietary company relationships dataset and linked, where possible, to a separate human-curated business conduct risk database that serves as the source of incident labels. Of the 11{,}373{,}899 entities in the ownership graph, 297{,}642 are covered by this human-curated database and are therefore eligible to receive incident labels; the remainder contribute only relational context. Node attributes include company characteristics, historical incident information, and observability related indicators available at the prediction cutoff.


The relation set $\mathcal{R}$ includes subsidiary, investment, fund, historical ownership, and related corporate links. Relation type and direction were retained so that different inter-firm channels are able to contribute differently to the predictions.

\subsection{Hub Structure}
\label{sec:hub-structure}

To determine an appropriate receptive field, we measured the proportion of unlabeled entities reachable from at least one labeled-positive entity within $k$ graph hops. Table~\ref{tab:hop-reachability} compares reachability with and without paths that traverse hubs, defined as entities with at least 1,000 neighbors.

\begin{table}[t]
\centering
\caption{Percentage of unlabeled entities reachable from at least one labeled-positive entity within $k$ hops, with and without hub-mediated paths.}
\label{tab:hop-reachability}
\begin{tabular}{c cc}
\toprule
& \multicolumn{2}{c}{\textbf{Reachable entities}} \\
\cmidrule(lr){2-3}
\textbf{Hops} & \textbf{With hubs} & \textbf{Without hubs} \\
\midrule
1 & 9.4\% & 8.8\% \\
2 & 22.5\% & 21.2\% \\
3 & 92.4\% & 26.4\% \\
4 & 93.7\% & 27.9\% \\
5 & 96.0\% & 28.3\% \\
6 & 96.1\% & 28.5\% \\
7+ & 96.2\% & 28.5\% \\
\bottomrule
\end{tabular}
\end{table}

With hub-mediated paths, reachability rises from 22.5\% at two hops to 92.4\% at three hops. Without hubs, three-hop reachability is only 26.4\% and plateaus at 28.5\%. The graph contains 230 entities with at least 1,000 neighbors and a further 8,700 with at least 100 neighbors. Median neighborhood size increases from 6 nodes at one hop to 122 at two hops and 3,029 at three hops.

These results motivate a three-hop receptive field: it captures most available connectivity, and deeper expansion would provide little additional reach. Because unrestricted traversal through hubs would create prohibitively large computation graphs, the model retains hubs but bounds neighborhood sampling, as described in Section~\ref{sec:subgraph-construction}.

\section{Methodology}
\label{sec:methodology}

\subsection{Problem Formulation and Leakage Control}
\label{sec:problem-formulation}

Let $G=(V,\{E_r\}_{r\in\mathcal{R}},X)$ denote the corporate relationship graph, where $V$ is the set of legal entities, $E_r$ is the set of directed edges associated with relation type $r\in\mathcal{R}$, and $X\in\mathbb{R}^{|V|\times d}$ contains node-level covariates available at the prediction cutoff. Each recorded company relationship and its reverse are represented as distinct relation types. The model can therefore learn different propagation patterns for, for example, parent-to-subsidiary and subsidiary-to-parent paths.

For each entity $v\in V$, the reference period supervision indicator is
\begin{equation}
s_v=
\begin{cases}
1, & \text{if $v$ has at least one recorded incident in 2024},\\
\varnothing, & \text{otherwise}.
\end{cases}
\end{equation}
The labeled positive and unlabeled sets are
\begin{equation}
P=\{v\in V:s_v=1\},
\qquad
U=V\setminus P.
\end{equation}
Membership in $U$ indicates only that no incident was recorded for the entity during the reference period; it does not establish that the entity is a true negative.

The model produces an unnormalized score $f_\theta(v)\in\mathbb{R}$ for each target entity. This score is used both to rank entities and compute the binary logistic loss. The forward evaluation outcome is
\begin{equation}
y_v^{\mathrm{future}}
=
\mathbf{1}\{\text{$v$ has at least one recorded incident in 2025}\}.
\end{equation}
The 2025 outcomes were not used in the model training, feature construction, hyperparameter selection, or model selection.

To prevent direct label leakage, the target entity's own 2024 supervision indicator is excluded from its input representation and used only in the training objective. Covariates available at or before the prediction cutoff and reference period information associated with neighboring entities remained available as relational context.

\subsection{Sampled Computation Graphs}
\label{sec:subgraph-construction}

Full neighborhood propagation is impractical in the graph described in Section 3.1 because a small number of corporate and fund entities have extremely large neighborhoods, causing rapid expansion of the computation graph across multiple layers. To bound memory and computation while retaining these high-degree entities, we construct target-centered computation graphs using fixed fanout neighborhood sampling, following the sampling and aggregation framework of GraphSAGE \citep{hamilton_inductive_2017}.

For a target entity $v$, let $V_v^{(0)}=\{v\}$. At depth $\ell\in\{1,\ldots,L\}$, the sampled node set is
\begin{equation}
V_v^{(\ell)}
=
V_v^{(\ell-1)}
\cup
\bigcup_{u\in V_v^{(\ell-1)}}
\operatorname{Sample}_{K}\!\left(\mathcal{N}(u)\right),
\end{equation}
where $\mathcal{N}(u)$ is the combined neighborhood of $u$ across all relation types and directions. The operator $\operatorname{Sample}_{K}$ retains all neighbors when $|\mathcal{N}(u)|\leq K$ and otherwise selects $K$ neighbors uniformly without replacement. We use a maximum depth of $L=3$ and fanout $K=100$. Relation types and edge directions are preserved after sampling.

This procedure does not exclude high-degree entities, but it limits the number of their neighbors included in each target specific computation graph. It therefore bounds memory use and computational cost while preserving sampled paths through hubs. The choice of a three-hop receptive field is informed by the topology analysis in Section~\ref{sec:hub-structure}, where reachability increases sharply between two and three hops and changes only marginally at greater depths. Because only a subset of neighbors is retained around high-degree entities, the sampled computation graphs do not necessarily preserve the full graph reachability reported in that analysis.

\subsection{Relation-Aware GCNII Encoder}
\label{sec:hetero-gcnii}

We encode each sampled computation graph using a heterogeneous extension of GCNII, which we call \emph{HeteroGCNII}. Standard graph convolution aggregates normalized information from adjacent nodes \citep{kipf_semisupervised_2017}. GCNII augments this operation with an initial residual connection and an identity mapping, which helps preserve node specific information and reduce over-smoothing across multiple graph convolution layers \citep{chen_simple_2020}. To model the directed, multi-relational ownership graph, HeteroGCNII applies an independent GCNII propagation channel to each directed relation type, following the general principle of relational graph convolution \citep{schlichtkrull_modeling_2018}.

The input features are first projected into a shared hidden representation:
\begin{equation}
H^{(0)}
=
\sigma\!\left(XW_{\mathrm{in}}+b_{\mathrm{in}}\right),
\end{equation}
where $W_{\mathrm{in}}$ and $b_{\mathrm{in}}$ are trainable parameters and $\sigma$ denotes the ReLU activation.

At graph convolution layer $\ell$, each directed relation type $r\in\mathcal{R}$ is processed by an independent GCNII propagation operator:
\begin{equation}
M_r^{(\ell)}
=
\mathcal{G}_{r}^{(\ell)}
\!\left(H^{(\ell)},H^{(0)},E_r\right).
\end{equation}

The relation specific representations are aggregated by summation:
\begin{equation}
H^{(\ell+1)}
=
\sigma\!\left(
\sum_{r\in\mathcal{R}}M_r^{(\ell)}
\right).
\end{equation}
Each relation channel has its own trainable parameters, while the initial representation $H^{(0)}$ is shared across channels. Because forward and reverse company relationships are represented as separate elements of $\mathcal{R}$, the model can learn asymmetric contributions for opposite directions of the same underlying relationship.

After $L$ layers, the score for target entity $v$ is
\begin{equation}
f_\theta(v)
=
w_{\mathrm{out}}^\top h_v^{(L)}+b_{\mathrm{out}},
\end{equation}
where $h_v^{(L)}$ is the final representation of $v$. The architecture and optimization hyperparameters are selected using the validation procedure described in Section 4.

\subsection{Non-Negative PU (nnPU) Risk Estimation}
\label{sec:nnpu}

A conventional binary objective would assign label 0 to every entity in $U$. This is inappropriate because the unlabeled population may contain firms with unobserved positive outcomes. We therefore train the model using the non-negative PU (nnPU) risk estimator proposed by \citet{kiryo_positive_2017}.

Let $\ell(f_\theta(v),y)$ denote the binary logistic loss for score $f_\theta(v)$ and label $y\in\{0,1\}$. For labeled positive and unlabeled minibatch sets $P_b$ and $U_b$, define
\begin{align}
\widehat{R}_{p}^{+}
&=
\frac{1}{|P_b|}
\sum_{v\in P_b}
\ell(f_\theta(v),1),\\
\widehat{R}_{p}^{-}
&=
\frac{1}{|P_b|}
\sum_{v\in P_b}
\ell(f_\theta(v),0),\\
\widehat{R}_{u}^{-}
&=
\frac{1}{|U_b|}
\sum_{v\in U_b}
\ell(f_\theta(v),0).
\end{align}
Given the latent positive class prior $\pi_p$, the empirical negative class risk is
\begin{equation}
\widehat{R}_{\mathrm{neg}}
=
\widehat{R}_{u}^{-}
-
\pi_p\widehat{R}_{p}^{-}.
\end{equation}
The subtraction corrects for the expected contribution of positive examples contained in the unlabeled population. The final nnPU objective is
\begin{equation}
\widehat{R}_{\mathrm{nnPU}}
=
\begin{cases}
\pi_p\widehat{R}_{p}^{+}+\widehat{R}_{\mathrm{neg}},
&
\widehat{R}_{\mathrm{neg}}\geq-\beta,\\
-\gamma\widehat{R}_{\mathrm{neg}},
&
\widehat{R}_{\mathrm{neg}}<-\beta.
\end{cases}
\end{equation}

We used $\pi_p=0.12$, $\beta=0$, and $\gamma=1$ in the reported experiments. The positive class prior was a domain-informed estimate based on risk probabilities observed in a broader proprietary risk dataset, which provides wider risk coverage than the human-curated dataset used to construct the study labels. The human-curated dataset was used for supervision because of its higher label quality, whereas the broader dataset was used only to inform the expected prevalence of underlying risk. These broader coverage estimates suggest that the latent positive prevalence was at least 12\%. We therefore used $0.12$ as a conservative operational prior.



The non-negativity correction prevents a flexible model from driving the empirical negative risk below zero and overfitting the labeled positive sample. Unlike cross-entropy training, the nnPU objective does not require unlabeled entities to be interpreted as confirmed negatives.

The nnPU estimator corrects for positive contamination in the unlabeled set but assumes that the labeled positive sample is sufficiently representative of the latent positive population. Selective incident visibility may violate this assumption, and the externally specified class prior is itself uncertain. The method therefore does not eliminate observability bias. We addressed this limitation through forward-looking evaluation and performance analyses stratified by historical visibility, company size, geography, and sector.

\section{Experiments and Results}
\label{sec:results}

\subsection{Experimental Design}
\label{sec:experimental-design}

\begin{table}[t]
\centering
\caption{Summary of the corporate ownership graph and temporal modelling cohort.}
\label{tab:dataset-summary}
\begin{tabular}{lr}
\hline
\textbf{Quantity} & \textbf{Value} \\
\hline
Entities in the modelling graph, $|V|$ & 11{,}373{,}899 \\
\quad of which are in the human-curated risk database & 297{,}642 \\
Directed ownership relations, $\sum_{r\in\mathcal{R}}|E_r|$ & 14{,}100{,}306 \\
Ownership-relation types, $|\mathcal{R}|$ & 8 \\
Node-feature dimensions, $d$ & 7 \\
Reference-period entities (2024) & 297{,}642 \\
Labeled-positive entities, $|P|$ & 22{,}990 \\
Unlabeled entities, $|U|$ & 274{,}652 \\
Validation entities & 10{,}000 \\
Evaluation entities (2025) & 297{,}642 \\
Positive evaluation outcomes, $\sum_v y_v^{\mathrm{future}}$ & 26{,}509 \\
\hline
\end{tabular}
\end{table}

The empirical analysis used a prospective design. Data available through 2024 were used for model development, and incidents recorded during 2025 were reserved for evaluation. An entity was treated as labeled positive during model development if at least one conduct-related incident was recorded between 2024-01-01 and 2024-12-31; all remaining entities were treated as unlabeled. The evaluation outcome was whether at least one incident was recorded between 2025-01-01 and 2025-12-31. No 2025 outcomes were used in the feature construction, training, hyperparameter selection, or model selection. The target entity's own 2024 supervision indicator was excluded from its input representation, as described in Section~\ref{sec:problem-formulation}.

The resulting scores were evaluated as ranking signals for investigative prioritization rather than as direct estimates of misconduct. The objective was to place entities with subsequently recorded incidents near the top of the 2025 ranking, including entities with limited earlier incident histories.

Table~\ref{tab:dataset-summary} summarizes the corporate graph and temporal modelling cohort after entity matching and application of the study inclusion criteria. The graph contains 11{,}373{,}899 entities, 14{,}100{,}306 directed relations, and 8 relation types. The node-feature matrix contains seven dimensions derived from company characteristics, historical information available at the prediction cutoff, and observability-related attributes. The 2024 development cohort contains 22{,}990 labeled-positive entities and 274{,}652 unlabeled entities, and 26{,}509 entities have a positive evaluation outcome in 2025.

Seven ranking procedures were evaluated. Random ranking provides a chance-level reference. Random forest \citep{breiman_random_2001} uses the node feature matrix without graph relations. A neighbor count heuristic assesses whether node degree alone explains the predictive value of the graph. Label propagation \citep{zhu_semisupervised_2003} evaluates non-parametric diffusion over the graph topology. Labeled positive entities are assigned a seed value of 1 and unlabeled entities a seed value of 0, with the target entity's own supervision indicator masked during scoring.

Two neural graph specifications were used to assess the contribution of relation-aware propagation. The homogeneous specification applies GCNII after collapsing all relation specific edge sets into a single graph. The relation-aware specification retains relation types and directions through separate propagation channels. The latter was evaluated with both cross-entropy and nnPU training. The homogeneous and relation-aware nnPU results provide a matched comparison of graph representation, while the two relation-aware results provide a matched comparison of training objectives.

Performance was measured using Precision@10\%recall, AUC-PR, AUC-ROC, NDCG, NDCG@10\%data, and Lift@10\%data. Precision@10\%recall was used for model selection, and it reports precision at the threshold required to recover 10\% of entities with positive 2025 outcomes. AUC-PR is the principal threshold averaged measure under class imbalance, and AUC-ROC was reported for completeness. NDCG and NDCG@10\%data measure ranking quality over the complete ranking and the top 10\% of entities, respectively \citep{jarvelin_cumulated_2002}.

Lift@10\%data is the precision among the top 10\% of ranked entities divided by the positive prevalence in the evaluated population. A value of 3 therefore indicates that the top ranked decile contains three times the positive outcome prevalence expected under a random selection of the same size. For the two-way visibility analysis, lift is also reported at the Precision@10\%recall operating point. Lift@10\%data is emphasized in subgroup comparisons because it normalizes performance by the positive prevalence within each stratum.

Hyperparameters were selected on a held-out validation set of 10{,}000 entities sampled using random seed 42. The neural model search covered one to three graph layers, hidden dimensions in $\{4,8,16,32,64,128,256\}$, and learning rates in $\{5\times10^{-3},10^{-3},5\times10^{-4}\}$. Cross-entropy and nnPU objectives were considered where applicable. Label propagation settings were selected using the same validation criterion; the selected configuration used three propagation layers and $\alpha=0.9$.

The reported homogeneous and relation-aware GCNII specifications used three graph layers and hidden dimensions of 32 and 64, respectively. Both used a learning rate of $10^{-3}$, batch size of 64, and gradient accumulation factor of 1. Neural models were trained for at most two epochs, with early stopping often terminating training earlier. The nnPU parameters were fixed at $\pi_p=0.12$, $\beta=0$, and $\gamma=1$, as described in Section~\ref{sec:nnpu}.

\subsection{Overall and Stratified Performance}

\begin{table*}[t]
\centering
\caption{Model performance on the held-out 2025 evaluation set. Lift@10\%data is top-decile precision divided by the prevalence of $y_v^{\mathrm{future}}=1$ in the evaluation population. Best observed values are shown in bold. The final neural models used a learning rate of $10^{-3}$, batch size of 64, three graph layers, and a maximum of two training epochs with early stopping.}
\label{tab:model-performance}
\resizebox{\textwidth}{!}{%
\begin{tabular}{llcccccccc}
\hline
\textbf{Model} & \textbf{Loss} & \textbf{\#Layers} & \textbf{Hidden} & \textbf{Prec.@10\%recall} & \textbf{AUC-PR} & \textbf{AUC-ROC} & \textbf{NDCG} & \textbf{NDCG@10\%data} & \textbf{Lift@10\%data} \\
\hline
Random & -- & -- & -- & 0.091 & 0.089 & 0.499 & 0.785 & 0.091 & 1.02 \\
Non-graph Random Forest & -- & -- & -- & 0.139 & 0.099 & 0.553 & 0.800 & 0.159 & 1.63 \\
Neighbour count & -- & -- & -- & 0.218 & 0.122 & 0.606 & 0.804 & 0.230 & 2.35 \\
Label propagation & -- & 3 & -- & 0.336 & 0.194 & 0.698 & 0.838 & 0.332 & 3.39 \\
GCNII & nnPU & 3 & 32 & 0.377 & 0.201 & 0.711 & 0.864 & 0.319 & 3.07 \\
HeteroGCNII & CE & 3 & 64 & 0.459 & 0.234 & 0.723 & 0.873 & 0.358 & 3.49 \\
HeteroGCNII & nnPU & 3 & 64 & \textbf{0.529} & \textbf{0.243} & \textbf{0.732} & \textbf{0.876} & \textbf{0.360} & \textbf{3.51} \\
\hline
\end{tabular}%
}
\end{table*}

\noindent\textbf{RQ1: Does observed conduct risk coverage vary systematically across firms?}

Historical incident coverage varies substantially across firm characteristics. Chi-square tests rejected equal coverage across market cap categories, the selected country groups, and sectors (all $p<0.001$; Table~\ref{tab:rq3-subgroup}, panels b--d). The range of observation rates across these groups indicates that missing incident records are systematically associated with company size, geography, and sector rather than being uniformly distributed across the population.

\medskip
\noindent\textbf{RQ2: Do graph-based models outperform non-graph baselines in predicting future recorded incidents?}

Table~\ref{tab:model-performance} reports the performance on the held-out 2025 outcomes. All graph-based procedures outperformed random ranking, the feature-only random forest, and the neighbor count heuristic on the principal ranking measures. The highest observed values were obtained by the relation-aware GCNII specification trained with nnPU: its Precision@10\%recall is 0.529, AUC-PR is 0.243, and Lift@10\%data is 3.51. The corresponding random forest values are 0.139, 0.099, and 1.63.

Label propagation achieves a Lift@10\%data of 3.39, indicating that ownership topology alone contains substantial forward-looking information. Under the same nnPU objective, retaining relation types and directions increases the Precision@10\%recall from 0.377 to 0.529 and AUC-PR from 0.201 to 0.243 relative to the homogeneous graph specification. Holding the relation-aware encoder fixed, replacing cross-entropy with nnPU increases Precision@10\%recall from 0.459 to 0.529 and AUC-PR from 0.234 to 0.243. The remaining reported measures change in the same direction. Because uncertainty intervals are not reported, these comparisons should be interpreted as observed differences rather than statistically significant claims.

\noindent\textbf{RQ3: Where are the predictive gains from graph-based inference strongest?}

\begin{table*}[t]
\centering
\caption{Observability-stratified performance of HeteroGCNII with nnPU and the random baseline. Visibility was defined using incidents recorded before 2024. Lift values were normalized by the prevalence of $y_v^{\mathrm{future}}=1$ within each 2025 evaluation stratum.}
\label{tab:rq3-visibility}
\resizebox{\textwidth}{!}{%
\begin{tabular}{llccccccc}
\hline
\textbf{Model} & \textbf{Dataset} & \textbf{Prec.@10\%recall} & \textbf{AUC-PR} & \textbf{AUC-ROC} & \textbf{NDCG} & \textbf{NDCG@10\%data} & \textbf{Lift@10\%recall} & \textbf{Lift@10\%data} \\
\hline
\multirow{2}{*}{HeteroGCNII (nnPU)}
  & Visible ($\geq$1 incident before 2024)
    & 0.663 & 0.400 & 0.695 & 0.905 & 0.513 & 3.46 & 2.55 \\
  & Non-visible (0 incidents before 2024)
    & 0.102 & 0.067 & 0.656 & 0.739 & 0.218 & 3.02 & 2.37 \\
\hline
\multirow{2}{*}{Random baseline}
  & Visible ($\geq$1 incident before 2024)
    & 0.191 & 0.192 & 0.500 & 0.835 & 0.192 & 1.00 & 1.00 \\
  & Non-visible (0 incidents before 2024)
    & 0.034 & 0.034 & 0.501 & 0.691 & 0.089 & 1.00 & 1.00 \\
\hline
\end{tabular}%
}
\end{table*}

Table~\ref{tab:rq3-visibility} separates entities according to whether an incident was recorded before 2024. Table~\ref{tab:rq3-subgroup} reports additional results by visibility threshold, country, market cap range, and sector. Because positive prevalence differs across these populations, the comparisons are based primarily on Lift@10\%data.


The Lift@10\%data is 2.55 among entities with at least one recorded incident before 2024 and 2.37 among entities with no incident before 2024. Relational information therefore remains useful among companies that entered the reference year without a recorded incident history. In the cumulative visibility groups, lift declines to 1.47 among entities with at least 10 earlier incidents, a population with substantially higher positive prevalence.

Across countries, lower observation rates are generally associated with larger lift values. Ireland, the United Kingdom, and Germany have observation rates near 23\% and lifts of 3.89--4.57, whereas Russia, China, and Brazil have higher observation rates and lifts of 2.25--2.76. These values are descriptive and do not constitute a complete country ranking.

The association is less regular across market-cap groups. Firms with market capitalization below USD 10 million have an observation rate of 14.7\% and lift of 3.47. The largest reported group has an observation rate of 66.1\% and lift of 2.47, and the intermediate groups do not follow a monotonic pattern.

The sector results show the clearest descriptive relationship. ``Software and Computer Services,'' ``Automobiles and Parts,'' and ``Telecommunications'' combine relatively low observation rates with lifts of 4.921, 4.553, and 4.323, respectively. ``Airlines'' and ``Food and Beverage'' have higher observation rates and lower respective lifts of 1.890 and 2.720. Overall, the subgroup results are consistent with larger relative gains where direct incident coverage is weaker, although this pattern is not uniform across all strata.

\begin{table*}[t]
\centering
\caption{Subgroup statistics and relation-aware nnPU performance by visibility threshold (a), selected country (b), market-cap range (c), and sector (d). Count and Obs.\ Rate are full-sample descriptive statistics. Prec.@10\% and Lift were computed on the 2025 evaluation set using the ranking induced by $f_\theta(v)$.}
\label{tab:rq3-subgroup}
\begin{minipage}[t]{0.38\textwidth}
\centering
{\small (a) Visibility threshold}\\[4pt]
\resizebox{\linewidth}{!}{%
\begin{tabular}{l rr rr}
\hline
 & \multicolumn{2}{c}{\textit{All time}} & \multicolumn{2}{c}{\textit{Test set (2025)}} \\
\cmidrule(lr){2-3}\cmidrule(lr){4-5}
\textbf{Visibility group} & \textbf{Count} & \textbf{Obs.\ Rate} & \textbf{Prec.@10\%} & \textbf{Lift} \\
\hline
Non-visible (0 incidents before 2024) & 215{,}797 & 6.7\% & 0.080 & 2.37 \\
Low visible (1--2 incidents before 2024) & 51{,}730 & 100\% & 0.200 & 2.06 \\
High visible ($\geq$2 incidents before 2024) & 44{,}176 & 100\% & 0.609 & 2.13 \\
Very high visible ($\geq$10 incidents before 2024) & 8{,}841 & 100\% & 0.877 & 1.47 \\
All visible ($\geq$1 incident before 2024) & 81{,}845 & 100\% & 0.490 & 2.55 \\
\hline
\end{tabular}%
}
\end{minipage}%
\hfill
\begin{minipage}[t]{0.30\textwidth}
\centering
{\small (b) Country (selected)}\\[4pt]
\resizebox{\linewidth}{!}{%
\begin{tabular}{l rr rr}
\hline
 & \multicolumn{2}{c}{\textit{All time}} & \multicolumn{2}{c}{\textit{Test set (2025)}} \\
\cmidrule(lr){2-3}\cmidrule(lr){4-5}
\textbf{Country} & \textbf{Count} & \textbf{Obs.\ Rate} & \textbf{Prec.@10\%} & \textbf{Lift} \\
\hline
United States & 82{,}506 & 29.1\% & 0.211 & 3.78 \\
United Kingdom & 24{,}773 & 23.1\% & 0.209 & 4.09 \\
China & 17{,}301 & 51.9\% & 0.336 & 2.76 \\
Germany & 14{,}895 & 22.9\% & 0.220 & 3.89 \\
\hdashline
India & 10{,}371 & 31.5\% & 0.331 & 3.81 \\
Japan & 8{,}387 & 37.2\% & 0.356 & 3.37 \\
\hdashline
Brazil & 3{,}722 & 50.9\% & 0.359 & 2.69 \\
Russia & 2{,}764 & 74.0\% & 0.448 & 2.25 \\
Ireland & 1{,}892 & 22.8\% & 0.189 & 4.57 \\
\hline
\end{tabular}%
}
\end{minipage}%
\hfill
\begin{minipage}[t]{0.28\textwidth}
\centering
{\small (c) Market cap (USD million)}\\[4pt]
\resizebox{\linewidth}{!}{%
\begin{tabular}{l rr rr}
\hline
 & \multicolumn{2}{c}{\textit{All time}} & \multicolumn{2}{c}{\textit{Test set (2025)}} \\
\cmidrule(lr){2-3}\cmidrule(lr){4-5}
\textbf{MCAP range} & \textbf{Count} & \textbf{Obs.\ Rate} & \textbf{Prec.@10\%} & \textbf{Lift} \\
\hline
$0 < \text{MCAP} \leq 10$ & 6{,}110 & 14.7\% & 0.088 & 3.47 \\
$10 < \text{MCAP} \leq 100$ & 13{,}312 & 21.4\% & 0.133 & 2.57 \\
$100 < \text{MCAP} \leq 1{,}000$ & 13{,}156 & 39.7\% & 0.383 & 3.12 \\
$1{,}000 < \text{MCAP} \leq 10{,}000$ & 5{,}848 & 66.1\% & 0.773 & 2.47 \\
\hline
\end{tabular}%
}
\end{minipage}

\vspace{8pt}

\par\noindent
{\small (d) Sector}\\[4pt]
\resizebox{\textwidth}{!}{%
\begin{tabular}{l rr rr l l rr rr l l rr rr}
\hline
 & \multicolumn{2}{c}{\textit{All time}} & \multicolumn{2}{c}{\textit{2025}} & &
 & \multicolumn{2}{c}{\textit{All time}} & \multicolumn{2}{c}{\textit{2025}} & &
 & \multicolumn{2}{c}{\textit{All time}} & \multicolumn{2}{c}{\textit{2025}} \\
\cmidrule(lr){2-3}\cmidrule(lr){4-5}
\cmidrule(lr){8-9}\cmidrule(lr){10-11}
\cmidrule(lr){14-15}\cmidrule(lr){16-17}
\textbf{Sector} & \textbf{Count} & \textbf{Obs.\ Rate} & \textbf{Prec.} & \textbf{Lift} & &
\textbf{Sector} & \textbf{Count} & \textbf{Obs.\ Rate} & \textbf{Prec.} & \textbf{Lift} & &
\textbf{Sector} & \textbf{Count} & \textbf{Obs.\ Rate} & \textbf{Prec.} & \textbf{Lift} \\
\hline
Airlines & 579 & 76.3\% & 0.793 & 1.890 & &
Travel and Leisure & 6{,}045 & 46.2\% & 0.458 & 2.601 & &
Aerospace and Defense & 1{,}760 & 55.1\% & 0.576 & 2.697 \\
Food and Beverage & 13{,}729 & 56.3\% & 0.422 & 2.720 & &
Retail & 12{,}338 & 37.5\% & 0.438 & 2.999 & &
Media & 5{,}857 & 32.1\% & 0.308 & 3.243 \\
Alternative Energy & 2{,}918 & 28.4\% & 0.366 & 3.310 & &
Industrial Transportation & 7{,}668 & 42.4\% & 0.464 & 3.338 & &
Construction and Materials & 16{,}122 & 39.2\% & 0.401 & 3.378 \\
Pharma.\ and Biotechnology & 7{,}561 & 35.7\% & 0.317 & 3.411 & &
Insurance & 4{,}899 & 29.8\% & 0.477 & 3.464 & &
Banks & 7{,}276 & 42.9\% & 0.489 & 3.482 \\
Industrial Metals & 4{,}916 & 39.3\% & 0.352 & 3.491 & &
Personal and Household Goods & 17{,}336 & 27.5\% & 0.241 & 3.498 & &
Mining & 9{,}572 & 35.1\% & 0.371 & 3.559 \\
Electronic and Electrical Equip. & 7{,}128 & 23.2\% & 0.198 & 3.615 & &
Health Care Equip.\ and Services & 12{,}353 & 22.0\% & 0.255 & 3.649 & &
Support Services & 24{,}701 & 23.6\% & 0.217 & 3.702 \\
Chemicals & 8{,}769 & 38.8\% & 0.303 & 3.707 & &
Technology Hardware and Equip. & 4{,}880 & 21.1\% & 0.196 & 3.742 & &
Utilities & 11{,}045 & 33.2\% & 0.447 & 3.760 \\
Oil and Gas & 9{,}596 & 40.3\% & 0.412 & 3.950 & &
Financial Services & 41{,}282 & 24.0\% & 0.277 & 3.955 & &
Industrial Engineering & 9{,}755 & 25.8\% & 0.242 & 4.240 \\
Telecommunications & 4{,}722 & 22.2\% & 0.319 & 4.323 & &
General Industrials & 4{,}159 & 26.4\% & 0.316 & 4.467 & &
Automobiles and Parts & 9{,}010 & 21.7\% & 0.319 & 4.553 \\
Software and Computer Services & 28{,}163 & 10.3\% & 0.159 & 4.921 & &
& & & & & &
& & & & \\
\hline
\end{tabular}%
}
\vspace{4pt}
\begin{minipage}{\textwidth}
\footnotesize
\raggedright
\textit{Notes:} Count and Obs.\ Rate are full-sample statistics covering the all-time incident history since 2007. Prec.@10\% and Lift are evaluated on the 2025 test set using $y_v^{\mathrm{future}}$. Chi-square tests reject equal historical incident coverage across countries (panel~b), market-cap categories (panel~c), and sectors (panel~d), all $p<0.001$. The visibility groups in panel~(a) are threshold-based and overlap by construction; for example, entities with two incidents appear in both the 1--2 and $\geq2$ groups. The all-time observation rate is descriptive and is not used as a future-dated model feature.
\end{minipage}
\end{table*}

\subsection{Relation-Level Analysis and Previously Non-Visible Firms}

\begin{table}[t]
\centering
\caption{Relation-level ablation results for the relation-aware nnPU specification. Importance is measured as the reduction in Lift@10\%data after partially or fully removing the relation-specific edge set $E_r$ from the test graph. The full-model baseline lift is 3.51.}
\label{tab:edge-importance}
\resizebox{\columnwidth}{!}{%
\begin{tabular}{l r cc}
\hline
& & \multicolumn{2}{c}{\textbf{Drop in Lift@10\%data}} \\
\cmidrule(lr){3-4}
\textbf{Relation type, $r$} & \textbf{$|E_r|$} & \textbf{500k removed} & \textbf{All removed} \\
\hline
Current Subsidiary/Operating Unit & 1{,}581{,}532 & 0.143 & 0.524 \\
Current Investment & 1{,}511{,}723 & 0.120 & 0.230 \\
Current Fund Sponsor & 7{,}955{,}990 & 0.082 & 0.134 \\
Prior Subsidiary/Operating Unit & 736{,}087 & 0.096 & 0.171 \\
Prior Investment & 967{,}619 & 0.026 & 0.134 \\
Other & 761{,}421 & 0.035 & 0.101 \\
Merged Entity & 353{,}798 & \multicolumn{1}{c}{---} & 0.196 \\
Current Fund Investment Advisor & 232{,}136 & \multicolumn{1}{c}{---} & 0.196 \\
\hline
\end{tabular}%
}
\vspace{1mm}
\begin{minipage}{\columnwidth}
\footnotesize
\raggedright
\textit{Notes:} Relations containing fewer than 500,000 edges do not have a 500k removal result. The 500k removal column provides a fixed volume comparison across sufficiently large relation sets, while the full removal column measures total contribution regardless of relation size.
\end{minipage}
\end{table}

\noindent\textbf{RQ4: Which ownership relation types contribute most to graph-based prediction?}

\begin{figure}[t]
\centering
\includegraphics[width=\columnwidth]{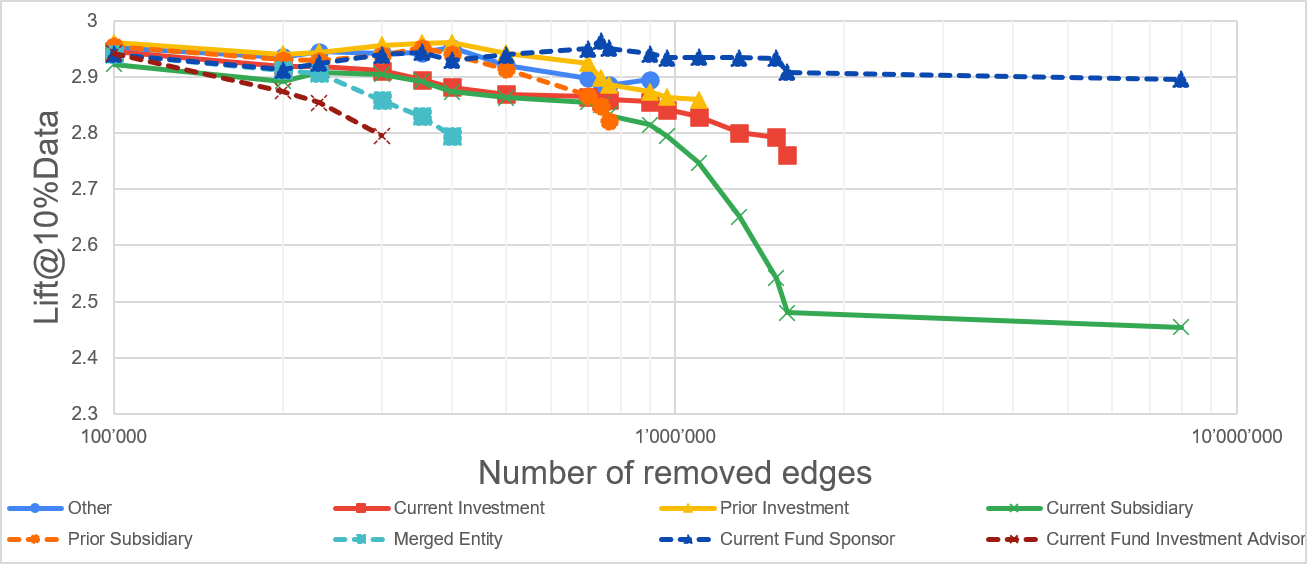}
\caption{Lift@10\%data after removing increasing numbers of edges from each relation-specific edge set $E_r$. The Lift@10\%data of the full model baseline is 3.51. Curves were smoothed using a three-point rolling average over adjacent ablation checkpoints for visual clarity. The results for Current Subsidiary/Operating Unit show the steepest sustained decline, whereas the results for Current Fund Sponsor remain comparatively stable until high removal volumes.}
\label{fig:edge-ablation}
\Description{Line chart showing Lift@10\%data as increasing numbers of edges are removed from each company relation type. Current subsidiary edges produced the largest sustained decline.}
\end{figure}

Relation-specific edge sets were ablated from the test graph and Lift@10\%data was recomputed after each removal step. Figure~\ref{fig:edge-ablation} reports the resulting trajectories. Table~\ref{tab:edge-importance} summarizes the reduction in lift after removing 500,000 edges (where the relation set is sufficiently large) and after removing the complete relation specific edge set.

The largest reduction follows the removal of the Current Subsidiary/Operating Unit edges. Removing all 1.58 million edges in this relation set lowers Lift@10\%data by 0.524 relative to the full model value of 3.51. Current Investment produces the second largest total reduction, 0.230. Prior Subsidiary/Operating Unit, Merged Entity, and Current Fund Investment Advisor show intermediate effects. By contrast, Current Fund Sponsor contains approximately 7.96 million edges but produces a total reduction of 0.134. Relation volume alone therefore does not explain predictive contribution.

\begin{figure}[t]
\centering
\includegraphics[width=0.9\columnwidth]{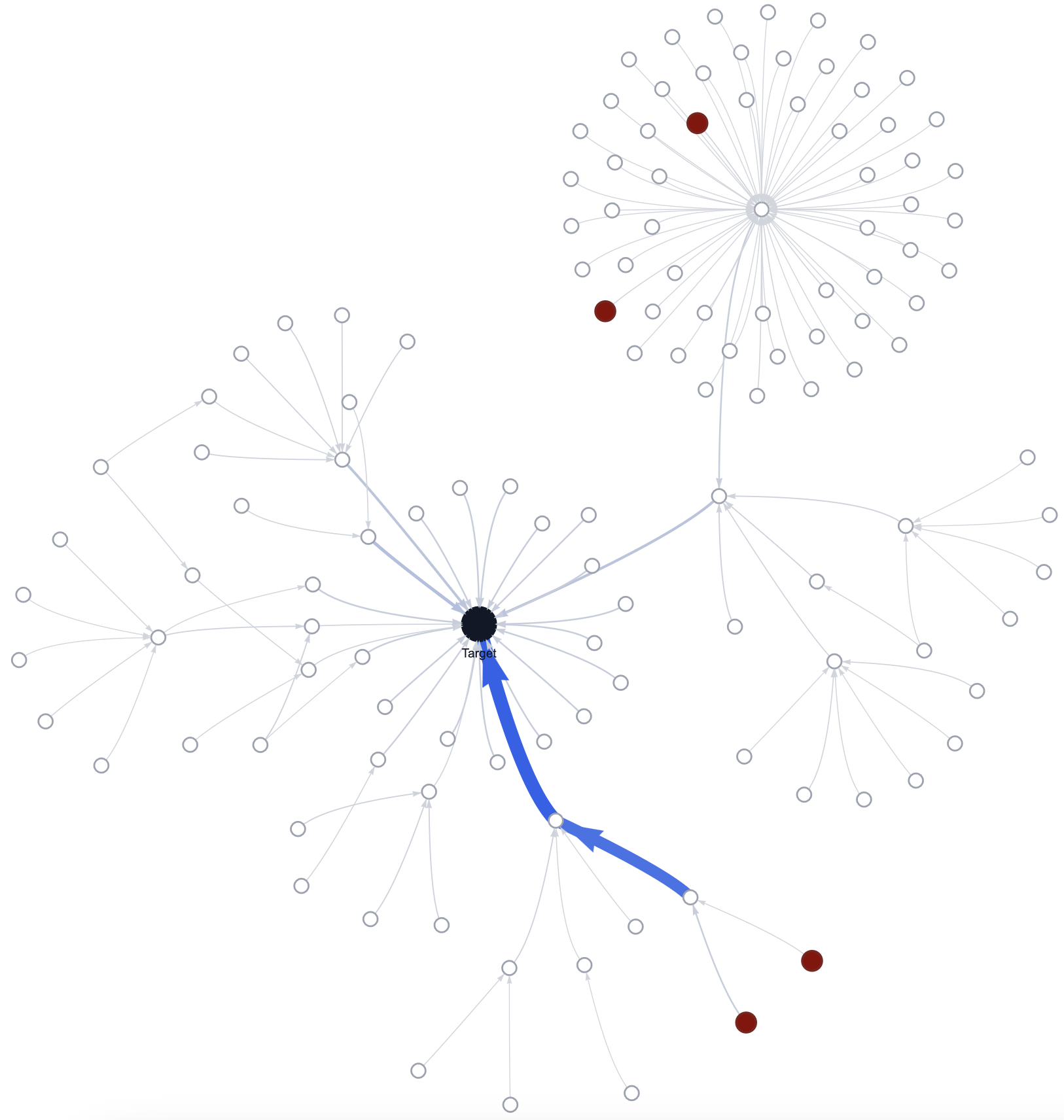}
\caption{Illustrative local explanation for a representative target entity (central black node).
Edge thickness indicates gradient based attribution, while node 
shading indicates known risk. Risk appears to propagate more strongly from the bottom nodes due to more impactful relationship types (e.g., 
Current Subsidiary at the bottom vs. Current Investment at the 
top). Aggregate relation-level importance is reported in Table~\ref{tab:edge-importance}. Additionally, risk from 
the top nodes is diluted across a massive hub with many non-risky 
connections, unlike the highly focused path at the bottom.}
\label{fig:explainer}
\Description{A local ownership subgraph in which thicker edges have larger gradient based attribution values and node shading indicates known risk. The lower relation path has a stronger contribution to the central target than the upper path through a large hub.}
\end{figure}

Figure~\ref{fig:explainer} provides an illustrative target-level explanation of this phenomenon. Edge thickness represents gradient based attribution, whereas node shading indicates known risk. In this example, the lower ownership path has a stronger contribution to the target prediction than the upper path. This difference is consistent with the more influential relation type in the lower path and with the dilution of information across the large hub in the upper path. Note that aggregate conclusions about relation importance are based on the ablation results rather than this single example.

\noindent\textbf{RQ5: Does graph-based inference retain predictive value among firms with no previously recorded incidents?}


Among the entities with no recorded incident before 2024, Lift@10\%data is 2.37 (Table~\ref{tab:rq3-visibility}). The ranking therefore retains its predictive value among firms that entered the reference year without an incident history, although some of these firms may become labeled positives during 2024.

\begin{figure}[t]
\centering
\includegraphics[width=0.9\columnwidth]{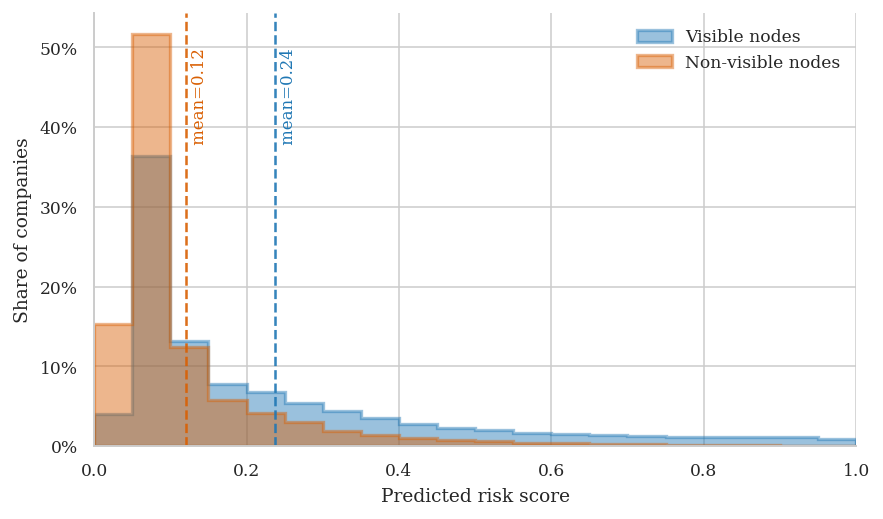}
\caption{Distribution of model scores $f_\theta(v)$ by visibility status before 2025. Visible entities have at least one recorded incident before 2025, while non-visible entities have none through 2024. Visible entities have a heavier right tail, while some non-visible entities also receive elevated scores.}
\label{fig:risk-distribution}
\Description{Overlaid score distributions for historically visible and non-visible entities. Both distributions concentrate near zero, while visible entities have a heavier right tail and some non-visible entities receive elevated scores.}
\end{figure}

Figure~\ref{fig:risk-distribution} applies a definition of visibility that is stricter than that used in Table 4 for the prediction cutoff: visible entities must have at least one incident recorded before 2025, whereas non-visible entities must have none through the end of 2024. Both score distributions are concentrated at low values, reflecting the overall class imbalance. The visible group has a heavier right tail, but elevated scores also occur among non-visible entities. The ranking therefore does not reduce to a prior-incident indicator. Because Table 4 and Figure 4 use different visibility cutoff dates, this figure is descriptive and does not provide a separate estimate of 2025 lift for the stricter no-incident-through-2024 group.


These results show that the relational structure can support prioritization among firms with limited prior visibility. They do not imply that every high scoring unlabeled entity is a hidden positive or that a low score establishes the absence of underlying conduct risk.

\section{Conclusion}

This paper considered the task of business conduct risk prediction, where the absence of a recorded incident could reflect limited observability rather than low underlying risk. We formulated the task as PU node classification on a directed corporate ownership graph and introduced HeteroGCNII, a relation-aware graph model trained with an nnPU objective.

In a forward-looking evaluation, the graph-based models outperformed feature-only baselines, and HeteroGCNII with nnPU achieved the strongest observed ranking performance. The model also provided a positive lift for firms that entered the reference period without a recorded incident history. Relation ablation identified current subsidiary and operating unit links as the most influential ownership channel, showing that corporate structure provides information beyond firm attributes and observed incident histories.


\section*{Acknowledgements}
R.H. is supported by the JST FOREST Program (JPMJFR216Q) and JSPS KAKENHI Grant Number JP24H00703.



\bibliographystyle{ACM-Reference-Format}
\bibliography{zotero_TIw,zotero_rh}

\end{document}